# BIOMECHANICS APPLIED TO COMPUTER-AIDED DIAGNOSIS: EXAMPLES OF ORBITAL AND MAXILLOFACIAL SURGERIES

Payan Yohan*, Luboz Vincent*, Chabanas Matthieu*, Swider Pascal**,
Marecaux C.*** & Boutault F.***

\* TIMC Laboratory, UMR CNRS 5525, University J. Fourier, 38706 La Tronche, France
\*\* Biomechanics Laboratory, EA 3697, Purpan University Hospital, 31059 Toulouse, France
\*\*\* Plastic and Maxillofacial Department, Purpan Univ. Hospital, 31059 Toulouse, France

RUNNING TITLE: Orbital and Maxillofacial Computer Assisted Surgeries

Corresponding authors:
Payan Yohan
Laboratoire TIMC/IMAG,
CNRS UMR 5525,
Institut d'Ingénierie de l'Information de Santé
Pavillon Taillefer – Faculté de Médecine
38706 La Tronche
France

Tel: +33 4 56 52 00 01
Fax: +33 4 56 52 00 55
e-mail: yohan.payan@imag.fr

**ABSTRACT**

This paper introduces the methodology proposed by our group to model the biological soft tissues deformations and to couple these models with Computer-Assisted Surgical (CAS) applications. After designing CAS protocols that mainly focused on bony structures, the Computer Aided Medical Imaging group of Laboratory TIMC (CNRS, France) now tries to take into account the behaviour of soft tissues in the CAS context. For this, a methodology, originally published under the name of the Mesh-Matching method, has been proposed to elaborate patient specific models. Starting from an elaborate manually-built "generic" Finite Element (FE) model of a given anatomical structure, models adapted to the geometries of each new patient ("patient specific" FE models) are automatically generated through a non-linear elastic registration algorithm.

This paper presents the general methodology of the Mesh-Matching method and illustrates this process with two clinical applications, namely the orbital and the maxillofacial computer-assisted surgeries.

## 1. Introduction

This paper aims at presenting the methodology proposed by our group to take into account the behaviour of biological soft tissues in the framework of Computer Aided Surgery (CAS). The CAS project, originally developed at the TIMC laboratory of Grenoble (CNRS, France) in the 80's, has proposed a lot of computer-aided clinical applications, most of them focusing onto orthopaedics. The spin-off company Praxim-Medivision (http://www.praxim.fr/), founded in 1995 by researchers from our group, is now commercializing these orthopaedic products.

Surgeries of bony structures were the first addressed by our group and by our industrial partners because bones are "quite easy" to track during surgery. Assuming that we are able to localize part of the structure (by fixing on it "Rigid Bodies" tracked in 3D inside the operating theatre), the complete geometry and position of the bone are known. It is therefore possible to assist the surgeon by providing him the actual position of the bone (see for example the CAS navigation system for the correct placement of pedicle screws [1]). This is unfortunately not possible for most organs and soft tissues that are supposed to move and to deform during surgery. In order to face this problem, researchers have tried to add a priori knowledge about the mechanical and/or physiological behaviour of such biological soft tissues, leading to CAS applications that try to model the tissues movements and deformations. Our group chose to develop biomechanical continuous models to predict the soft tissues deformations. These models are based on the Finite Element (FE) Method that discretizes the partial differential equations that govern the Continuum Mechanics. This paper aims at describing our methodology for soft tissue modelling and its coupling with CAS applications. The first part describes the complete methodology for building a patient-specific FE model from medical imaging exams (CT, MRI and/or US). The second part of the paper illustrates some clinical applications, namely the orbital and maxillofacial computer-aided surgeries.

## 2. Methodology for building patient-specific Finite Element models

2.1. Patient-specific FE meshes

Finite Element (FE) analysis is a widely used method in the field of biomechanics and customized meshes are of great interest since they can integrate both geometry and mechanical properties of the patient. However, except from very clear and normalized frameworks [2-3] or from the use of automatic un-structured tetrahedral mesh generators (available in almost any commercial FE package [4]), building a structured patient-specific FE model remains complicated and time consuming. Indeed, the mesh has to be adapted to the global patient geometry but usually needs to take into account some specific internal sub-structures, leading to topological changes (changes that allow, for the example of bones, to differentiate the cortical bone from the cancellous bone [5]). This organization of the FE mesh makes then possible to differentiate the sub-structures from a mechanical point of view, by assigning for example different Young modulus values to different elements inside the mesh [5].

The time consuming manual elaboration of such patient-specific FE meshes is unfortunately not always compatible with a clinical use. Moreover, specific exams such as thin inter-slices CT or MRI are needed to build an accurate FE mesh, but are not always available and used for each patient. To face these limits, our group proposed the Mesh-

Matching (MM) algorithm [6]. The idea is to start with a "generic mesh" of a given anatomical structure. This mesh is accurately designed, with a strong manual interaction leading to geometries and mesh topologies that are adapted to the structure: differentiation between sub-structures (for ex. cortical and cancellous bones), associations between organized (and labelled) elements and internal sub-structures such as muscles, dermis layers, etc… For a given anatomical structure, the researcher can spend hours or days designing this generic mesh. Once this work is done, the generic mesh is used in the clinical framework to automatically build patient-specific FE meshes. This is done through the following steps:

1. For each new patient, anatomical data (in general the external surface of the patient anatomical structure) are collected. This can be rough data (coming from US exam or from sparse CT/MRI exams) or fine data extracted through CT or MRI exams similar to those used to build the generic mesh. A set of 3D points located onto the external surface of the patient anatomical structure is therefore collected during this step.

2. An elastic registration method, originally proposed in the field of computer-assisted surgery [7], is used to match the extracted patient surface points with the nodes located on the external surface of the generic mesh. This matching aims at finding a volumetric transform **T**, which is a combination of global (rigid) and local (elastic) transforms. The idea underlying the matching algorithm consists (1) in aligning the two datasets (the rigid part of **T**) and (2) in finding local cubic B-Splines functions. The unknowns of the transform are all the B-Splines parameters. Those parameters are obtained through an optimization process that aims at minimizing the distance between the two surfaces, namely the points extracted from the patient data and the external nodes of the generic mesh.

3. The volumetric transform **T** is then applied to every node of the FE generic mesh, namely the nodes located on the external surface as well as the internal nodes that define the FE volume. A new volumetric mesh is thus automatically obtained by assembling the transformed nodes into elements, with a topology similar to that of the generic FE mesh: same number of elements and same element types.

4. The regularity [8] of the patient 3D mesh is checked in order to see if any FE analysis can be performed. If some elements of the mesh are detected as irregular, a global mesh regularization technique is proposed [9].

Figure 1 shows the results provided by the MM algorithm for the automatic generation of an entire patient FE femora mesh [10]. This example illustrates the MM methodology and can be straightforwardly applied to the case of soft tissues.

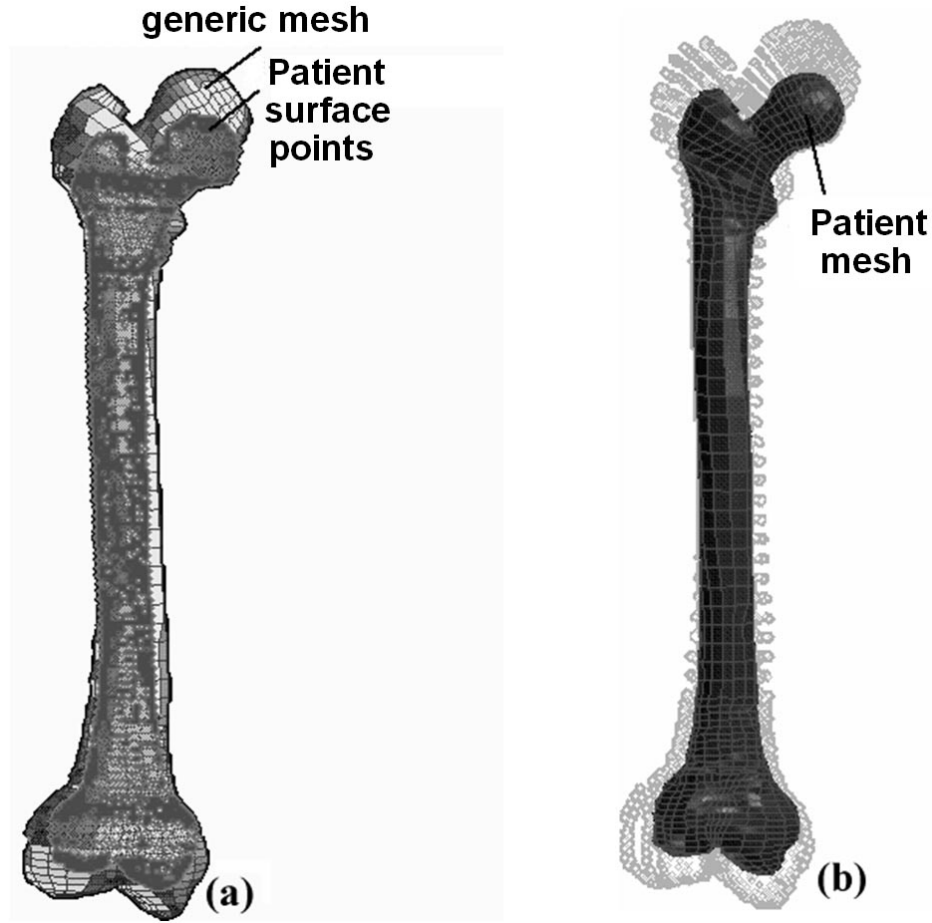

**Figure 1:** The Mesh-Matching algorithm applied to entire femora. (a) The external nodes of the generic mesh are matched onto patient surface points in order to compute a volumetric transform **T**. (b) **T** is then applied to all the generic nodes in order to generate the patient-specific FE mesh.

### 2.2. Geometrical and mechanical hypothesis for FE analysis

Once the patient specific mesh has been generated, hypotheses have to be made to model the soft tissues continuous mechanical behavior. Three different modeling hypotheses can be usually made to model soft tissues (see [11] for a review): (1) a linear elastic model assuming small deformations, (2) a linear elastic model under large deformation hypothesis, and (3) a hyperelastic model.

*2.2.1 Equations of the continuum mechanics*

Any object point is represented by its Lagrangian material coordinates $X = (X_1, X_2, X_3)$ at the undeformed state and its Eulerian spatial coordinates $x = (x_1, x_2, x_3)$ at the deformed state. The continuum mechanics introduces the deformation gradient tensor $F$, to relate the deformed and undeformed state (eq. 1), and the Lagrangian strain tensor $E$ (eq. 2) that is a more suitable measure of deformation since it reduces to the zero tensor for rigid-body motion.

$$F(\vec{X}) = \begin{bmatrix} \dfrac{\partial x_1}{\partial X_1} & \dfrac{\partial x_1}{\partial X_2} & \dfrac{\partial x_1}{\partial X_3} \\ \dfrac{\partial x_2}{\partial X_1} & \dfrac{\partial x_2}{\partial X_2} & \dfrac{\partial x_2}{\partial X_3} \\ \dfrac{\partial x_3}{\partial X_1} & \dfrac{\partial x_3}{\partial X_2} & \dfrac{\partial x_3}{\partial X_3} \end{bmatrix} \quad (1)$$

$$E = \frac{1}{2}(F^T F - I) = \frac{1}{2}(\nabla \mathbf{U}^T + \nabla \mathbf{U} + \nabla \mathbf{U}^T \nabla \mathbf{U}) \quad (2)$$

where $I$ is the identity matrix and $U = x - X$ is the displacement vector field.

A second tensor, the Cauchy stress tensor $T$, is introduced in order to characterize and to model internal forces. This tensor is a 3x3 symmetric tensor that must satisfy, for a static simulation framework, the quasi-static equilibrium (momentum conservation):

$$\text{div } \mathbf{T} + f = 0 \quad (3)$$

where $f$ are the applied volumetric forces.

In order to determine the changes of kinematic variables when forces are applied or to determine the changes of stress when strain is induced, a third equation must be introduced, to link the strain tensor $E$ and the stress tensor $T$. From a mathematical point of view, it can be described as:

$$\mathbf{T} = \mathrm{f}(E) \quad (4)$$

where f is a function that can have different formulation according to modeling assumptions.

*2.2.2 Small deformation linear elastic hypothesis*

This modeling framework is the most simple (and therefore limited) but the most widespread among the literature. It is based on two assumptions.

The first modeling assumption, called the *small deformation hypothesis* or the *hypothesis of linear geometry*, assumes that the deformations of the material are "small" (usually, a threshold of 10% of deformation is given). As a consequence of this assumption, equation (2) can be reduced to its linear part, by neglecting the second order terms:

$$E = \frac{1}{2}(\nabla \mathbf{U}^T + \nabla \mathbf{U}) \quad (5)$$

A second modeling assumption, known as the *linear mechanical hypothesis*, can be done by assuming a mechanical linearity between the stress and the strain tensor. Equation (4) can therefore be reduced to:

$$\mathbf{T} = \mathrm{C} \cdot E \quad (6)$$

where C is a fourth order tensor (3x3x3x3) named the *elastic tensor*. It can be seen that under assumptions of homogeneity and isotropy, this tensor is characterized by only two coefficients, namely the Young modulus $E$ and the Poisson's ratio $v$. The Young modulus is a kind of measure of the material stiffness while the Poisson's ratio is correlated to the compressibility of the material.

Thanks to the modeling assumptions described above, the equations of the continuum mechanics can be reduced to matrix equations with direct solutions that can be numerically computed. This explains why the small deformation linear elastic assumption is the most commonly used among the literature.

*2.2.3 Large deformation linear elastic hypothesis*

This modeling framework is usually more adapted to biological soft tissues as it allows levels of deformations that undergo 10%. The idea consists in still assuming a linear mechanical hypothesis, but without neglecting the second order terms of equation (2). This modeling framework is commonly assumed to be more accurate than the small deformation linear elastic hypothesis for tissues that show strong deformations, but needs larger computation times as the solution is obtained through an iterative optimization technique.

*2.2.4 Hyperelastic hypothesis*

This modeling framework can be even more adapted to biological soft tissue as it allows, in addition to the large deformation framework, a constitutive law (stress/strain relationship) that is non-linear. Indeed, most of the soft tissue show an "exponential-like" behavior for the constitutive law [12], with a stiffening of the tissue when the deformations increase. The hypothesis of *hyperelasticity* tries to model this, by assuming that the stress *T* can be derived from the strain tensor *E* and from a stored strain energy function *W*:

$$\mathbf{T} = \frac{\partial W}{\partial E} \tag{7}$$

The strain energy *W* is a function of multidimensional interactions described by the nine components of *F*. It is very difficult to perform experiments to determine these interactions for any particular elastic material. Therefore, various assumptions have been made to derive simplified and realistic strain energy functions, and different formulations have been elaborated, such as the ones of the Ogden [13], the Yeoh [14] or the Mooney-Rivlin [15].

*2.2.5 Methodology for choosing the most appropriate hypothesis*

Our approach for choosing the most appropriate modeling hypothesis can be defined as "pragmatic". The idea is first to test and quantitatively evaluate the most simple hypotheses, namely the linear geometrical and/or mechanical hypotheses (see [16] for a detail evaluation of different modeling assumptions in the context of computer-aided maxillo-facial surgery). If it appears that a non-linear model is needed, an iterative method has been proposed to infer, from indentation experiments that measure in vivo or in vitro force/displacement relationship, the hyperelastic strain energy function (and therefore the constitutive law) of the material. This method is based on a Finite Element Analysis of the indentation experiment. An optimization process is used to determine the FE constitutive laws that provide the non-linear force/displacements observed during the indentation experiments (see [17] for a complete description of the method).

## 3. Orbital and maxillofacial computer-aided surgeries

*3.1 Exophtalmia*

Exophtalmia is characterized by a forward displacement of the eye ball outside the orbit due to a pathology that increases the volume of the ocular muscles and/or the orbital fat tissues [18]. The functional consequences are a too long cornea exposition or, in the worst case, a distension of the ocular nerve that can lead to a decrease of the visual acuity. One of the treatments for exophthalmia is surgery, with an osteotomy (i.e. a hole in the maxillary or ethmoid sinuses regions) of the orbital walls that aims at increasing the volume of the orbital cavity [19]. To improve the backward displacement of the eye ball, some surgeons push on it in order to evacuate more of the fat tissues in the sinuses.

Up to now, the predictions of the consequences of an exophthalmia reduction were based on clinical observations [20] that state that for a 1 $cm^3$ soft tissues decompression, a backward displacement from 1 mm to 1.5 mm is expected. In order to complete this experimental clinical rule, our group has proposed to build a numerical model of the surgical gesture [21]. For this, a FE poro-elastic model of the orbital content (muscles+fat tissues) is used to predict the globe backward displacement assuming (1) a given size and position of the osteotomy and (2) a given pressure exerted by the surgeon onto the globe. The MM methodology proposed by our group has been used here, with a generic FE mesh of the orbital content (figure 2.a) adapted to different patients morphologies (figure 2.b).

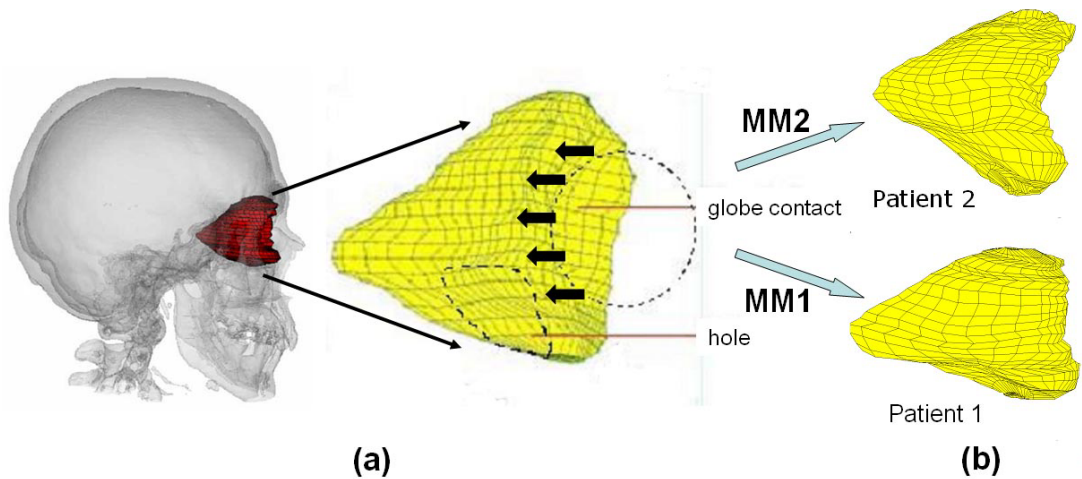

**Figure 2:** Poroelastic FE generic model of the orbital soft tissues (a) adapted to different patient morphologies with the Mesh-Matching algorithm (b)

For a given location of the osteotomy, a mean rule has been extrapolated from the numerical simulations provided by the model [22]: equation (8) links the suited globe backward displacement *disp* (mm) and the surface *surf* ($cm^2$) of the osteotomy:

$$disp = 1.1*\ln(surf)+1.9 \qquad (8)$$

*3.2 Orthognathic surgery*

Orthognathic surgery attempts to establish normal aesthetic and functional anatomy for patients suffering from dentofacial disharmony [23]. In this way, current surgery aims at normalize patients dental occlusion, temporo mandibular joint function and morphologic appearance by repositioning maxillary and mandibular skeletal osteotomized segments. The clinical predictions of the aesthetic and the functional consequences of the bone repositioning planning remain very qualitative. While some surgeons still work on patient front and profile photography and try to cut these photographs to qualitatively predict the patient face aesthetics after surgery, others try to use the computer framework to numerically simulate the mechanical behaviour of the facial soft tissues in response to the modification of the bone position [24-25]. Our group is working on this application since 2000, with a methodology that is still the same: (1) a generic FE model of the face soft tissue is built and (2) used to automatically generate patient specific models (figure 3) that (3) predict the surgical outcome (figure 4).

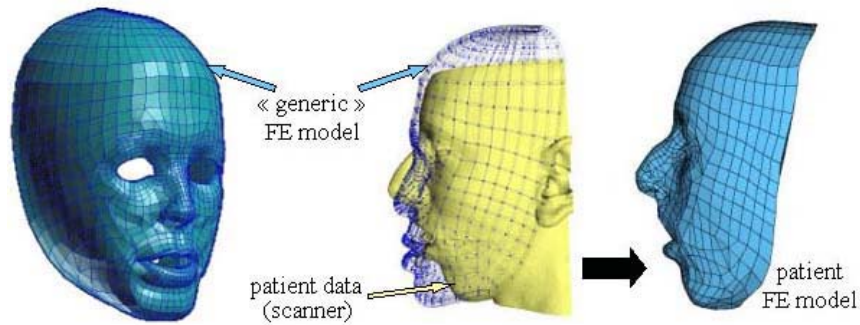

**Figure 3:** FE generic model of the face soft tissues adapted to a patient morphology with the Mesh-Matching algorithm.

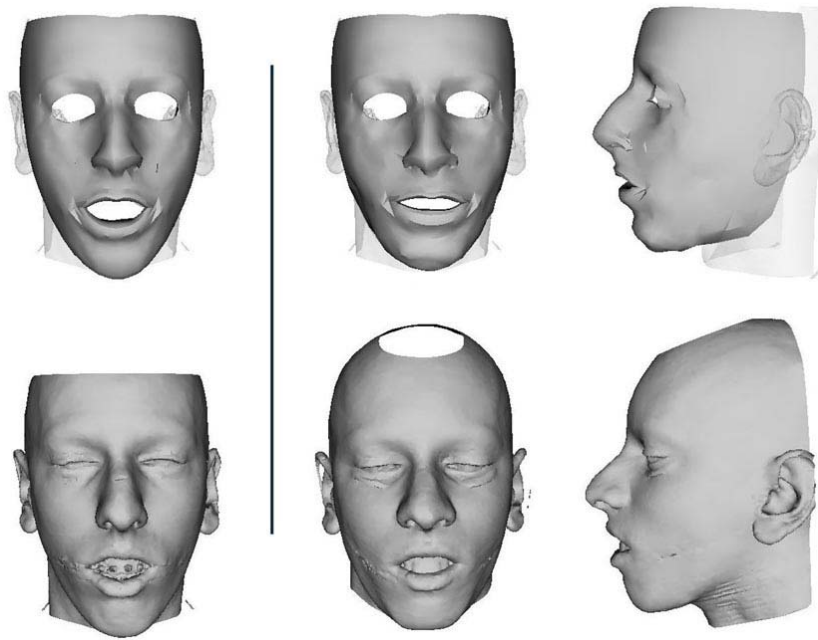

**Figure 3:** Qualitative evaluation. The simulations (top) are visually compared with the 3D reconstruction of the post-operative patient skin surface (bottom). Emphasis is given to the perception of the model quality in the most relevant morphological areas in the face: cheeks bones, lips area, chin and mandible angles.

The simulations provided by our face models were quantitatively evaluated with a patient for whom a post-operative CT exam was provided (see [16] for a complete description of this evaluation). For this case, we were able to measure the post-operative external surface of the patient face and to quantitatively compare this surface with the predictions provided by the models. Three types of models were evaluated for this study: a linear elastic model assuming small deformations, a linear elastic model under large deformation hypothesis, and a hyperelastic model. The surgical gesture consisted in a backward translation of 0.9 mm in the mandible axis and a slight rotation in the axial plane. The first interesting result was that the simulations obtained with all models were quite similar. More surprisingly, the results obtained with the hyperelastic model showed more important errors than the ones provided by the small deformation linear elastic model. We explained this by the fact that such hyperelastic modelling is much more sensitive to critera like the quality of the mesh, the boundary conditions and the rheological parameters. Such complicated models require more testing before being used, and may not be the most adapted for problems with relatively small deformations.

## 4. Conclusion

This paper aimed at introducing the methodology defined by our group to model biological soft tissues and to propose computer-assisted applications that integrate these models. A special focus was given to the compatibility of this approach with the clinical framework. Indeed, models need to be easily conformed to each new patient morphology, which is taken into account by the Mesh-Matching algorithm proposed by our group. A modelling approach, qualified as "pragmatic", was also depicted. The idea is that we do not think it is necessary to use the most complex mechanical modelling framework if this does not lead to an improvement in the accuracy of the numerical results compared with post-operative data. For this reason, we think a strong focus has to be done onto the validation of the results provided by the model, by quantitatively comparing the predictions with the surgical outcome.